\documentclass[prd,preprintnumbers,amsmath,amssymb,nofootinbib,showkeywords]{revtex4}
\usepackage{epsfig}
\usepackage{amssymb}
\usepackage{graphicx}
\usepackage{hyperref}
\usepackage[usenames]{color}

\def\eq#1{(\ref{#1})}

\def\s0#1#2{\mbox{\small{$ \frac{#1}{#2} $}}}
\def\0#1#2{\frac{#1}{#2}}

\def\beq{\begin{equation}}
\def\eeq{\end{equation}}

\def\bea{\arraycolsep .1em \begin{eqnarray}}
\def\eea{\end{eqnarray}}

\begin{document}
\title{Renormalisation group and the Planck scale}

\author{Daniel F. Litim}
\email{d.litim@sussex.ac.uk}
\affiliation{Department of Physics and Astronomy, University of Sussex, Brighton, BN1 9QH, U.K.}

\begin{abstract}
I discuss the renormalisation group approach to gravity, its link to S.~Weinberg's asymptotic safety scenario, and give an overview of results with applications to particle physics and cosmology. 
  \end{abstract}
\maketitle

%Preprint INT-PUB-11-004 

%********|*********|*********|*********|*********|*********|*********|****
\section{Introduction}
%********|*********|*********|*********|*********|*********|*********|****
Einstein's theory of general relativity is the remarkably successful classical theory of the gravitational force, characterised by Newton's coupling constant $G_N=6.67\times 10^{-11}{\rm m^3}/({\rm kg\,s^2})$ and a small cosmological constant $\Lambda$. 
Experimentally, its validity has been confirmed over many orders of magnitude in length scales ranging from the sub-millimeter regime up to solar system size. At larger length scales, the standard model of cosmology including dark matter and dark energy components fits the data well.
At shorter length scales, quantum effects are expected to become important. An order of magnitude estimate for the quantum scale of gravity  -- the Planck scale -- is obtained by dimensional analysis leading to the Planck length $\ell_{\rm Pl}\approx\sqrt{\hbar G_N/c^3}$ of the order of $10^{-33}\,{\rm cm}$, with $c$ the speed of light.  In particle physics units this translates into the Planck mass
\begin{eqnarray}
M_{\rm Pl}&\approx & 10^{19}\ {\rm GeV}\,.
\label{Planck}
\end{eqnarray}
While this energy scale is presently out of reach for earth-based particle accelerator experiments, fingerprints of Planck-scale physics can nevertheless become accessible through cosmological data from the very early universe.
From a theory perspective, it is widely expected that a fundamental understanding of Planck scale physics requires a quantum theory of gravity. It is well known that the standard perturbative quantisation programme faces problems, and  a fully satisfactory quantum theory, even outside the framework of local quantum physics, is presently not at hand.  
In the past 15 years, however, a significant body of work has been devoted to re-evaluate the physics of the Planck scale within conventional 
settings. Much of this renewed interest is fueled by Steven Weinberg's seminal proposal, more than 30 years of age,  that a quantum theory of gravitation may  {\it dynamically}  evade  the virulent divergences encountered in standard perturbation theory \cite{Weinberg}. This scenario, known as asymptotic safety, implies that gravity achieves a non-trivial ultraviolet (UV) fixed point under its renormalisation group flow. If so, this would incorporate gravity alongside the set of well-understood quantum field theories whose UV behaviour is governed by a fixed point, eg.~Yang-Mills theory. 

In this note,
I discuss the renormalisation group approach to gravity. I recall some of the issues of perturbative quantum gravity (Sec.~\ref{PT}), introduce the renormalisation group \`a la Wilson to access the physics at the Planck scale (Sec.~\ref{RG}), review key results in four dimensions (Sec.~\ref{FPs}), evaluate applications within low-scale quantum gravity (Sec.~\ref{Extra}), and conclude (Sec.~\ref{C}).

\section{Perturbative quantum gravity}\label{PT}
A vast body of work has been devoted to the perturbative quantisation programme of gravity.  I recall a very small selection of these in order to prepare for the subsequent discussion, and
I use particle physics units  $\hbar=c=k_B=1$ throughout.
Classical general relativity is described by the classical action
\begin{equation}\label{S}
S=\frac{1}{16\pi\,G_N}\int d^4x\sqrt{\det g_{\mu\nu}}\left(-R(g_{\mu\nu})+2\Lambda\right)
\end{equation}
where I have chosen a euclidean signature, $R$ denotes the Ricci scalar,  and $\Lambda$ the cosmological constant term.
The main point to be stressed is that the fundamental coupling of gravity $G_N$ in \eq{S} carries a dimension which sets a mass scale, with mass dimension $[G_N]=2-d$ in $d$ dimensional space-time. This structure distinguishes gravity in a profound manner from the other fundamentally known interactions in Nature, all of which have dimensionless coupling constants from the outset. Consequently, the effective {\it dimensionless} coupling of gravity which organises its perturbative expansion is given by 
\begin{equation}
\label{g}
g_{\rm eff}\equiv G_N\, E^2
\end{equation} 
(in four dimensions), where $E$ denotes the relevant energy scale. While the effective coupling \eq{g} remains small 
for energies below the Planck scale $E\ll M_{\rm Pl}$, it grows large in the Planckian regime where an expansion in $g_{\rm eff}$ may become questionable. Within the Feynman diagrammatic approach, this behaviour translates into the (with loop order) increasing degree of divergence of perturbative diagrams involving gravitons
\cite{Weinberg}. This structure is different from standard quantum field theories and relates to the classification of interactions as super-renormalisable, renormalisable, or `dangerous', depending on whether their canonical mass dimension is positive, vanishing, or negative. The degree of divergences implied by the negative mass dimension of Newton's coupling is mirrored in the perturbative non-renormalisability of Einstein gravity which has been established at the one-loop level  \cite{'tHooft:1974bx} in the presence of matter fields, and at the two-loop level  \cite{Goroff:1985sz} within pure gravity.

While this state of play looks discouraging from a field theory perspective, it does not rule out a quantum-field theoretical description of gravity altogether. There are a few indicators available to support this view.
Firstly, at low energies, a weak-coupling analysis of quantum gravity effects is possible within an effective theory approach \cite{Donoghue:1993eb}, 
which operates an ultraviolet cutoff at the Planck scale, see \cite{Burgess:2003jk} for a review. Secondly, higher order derivative operators appear to stabilise perturbation theory \cite{Niedermaier:2006ns}. Including all fourth order derivative operators it has been proven by K.~Stelle that gravity is  renormalisable to all orders in perturbation theory \cite{Stelle:1976gc}.  This striking difference with  Einstein-Hilbert gravity highlights the stabilising effect of higher derivative terms. Unfortunately, the resulting theory is not compatible with standard notions of perturbative unitarity and has therefore not been considered as a candidate for a fundamental theory of gravity. The r\^ole of higher order derivative operators has further been clarified in \cite{Gomis:1995jp} with the help of a BRST analysis. Interestingly, the theory remains unitary once all higher derivative operators are retained, whereas renormalisability is at best achieved in a very weak sense due to the required infinitely many counter terms. 

\section{Renormalisation group}\label{RG}

The renormalisation group (RG) comes into play when the running of couplings with energy is taken into account. As in any generic quantum field theory, quantum fluctuations modify the strength of couplings. If the metric field remains the fundamental carrier of the gravitational force, the fluctuations of space-time itself should modify the gravitational interactions with energy or distance. For Newton's coupling, this implies that $G_N$ becomes a running coupling $G_N\to G(k)=G_N\,Z^{-1}(k)$ as a function of the RG momentum scale $k$, where $Z(k)$ denotes the wave-function renormalisation factor of the graviton. Consequently, the dimensionless coupling $g_{\rm eff}$ in \eq{g} should be replaced by the running coupling
\begin{equation}\label{gk}
g=G(k)\, k^2
\end{equation}
which evolves with the RG scale. In particular, the UV behaviour of standard perturbation theory is significantly improved, provided that \eq{gk} remains finite in the high-energy limit. This is the asymptotic safety scenario as advocated by S.~Weinberg \cite{Weinberg}, see \cite{Niedermaier:2006ns,NiedermaierReuter} for extensive accounts of the scenario and \cite{Litim:2006dx,Percacci:2007sz,Litim:2008tt} for brief overviews. 
The intimate link between a fundamental definition of quantum field theory and  renormalisation group fixed points has been stressed by K.~Wilson some 40 years ago \cite{Wilson:1971bg,Wilson:1971dh}.
For gravity, the fixed point property becomes visible by considering the Callan-Symanzik-type RG equation for \eq{gk} which in $d$ dimensions takes the form \cite{Litim:2006dx} (see also \cite{Niedermaier:2006ns,NiedermaierReuter})
\begin{equation}\label{beta-g}
\partial_t g \equiv \beta_g = (d-2+\eta)\,g
\end{equation}
with $\eta=-\partial_t \ln Z(k)$ the graviton anomalous dimension which in general is a function of all couplings of the theory including matter, and $t=\ln k$. This simple structure arises provided the underlying effective action is local in the metric field.  From the RG equation \eq{beta-g} one concludes
that the gravitational coupling may display two types of fixed points. The non-interacting (Gaussian) fixed
point corresponds to $g_*=0$ and entails the vanishing of the anomalous dimension $\eta=0$.
In its vicinity, gravity stays classical, and $G(k)\approx G_N$.  On the other hand, a non-trivial RG fixed point with $g_*\neq 0$ 
can be achieved implicitly, provided that the anomalous dimension reads
\begin{equation}\label{eta}
\eta_*=2-d\,.
\end{equation}
The signficance of \eq{eta} is that the graviton anomalous dimension precisely
counter-balances the canonical dimension of Newton's coupling $G_N$. 
This pattern is known from other gauge systems at a critical point away from their 
canonical space-time dimensionality  \cite{Litim:2006dx}, eg. $U(1)$ Higgs theory in three dimensions \cite{Bergerhoff:1995zq}. 
In consequence, the 
dimensionful, renormalised coupling scales as $G(k)\approx g_*/k^{d-2}$
and becomes small in the ultraviolet limit where $1/k\to 0$. This pattern is at the
root for the non-perturbative renormalisability of quantum gravity within a
fixed point scenario. 

Much work has been devoted to check by explicit computation 
whether or not the gravitational  couplings achieve a non-trivial UV fixed 
point. A versatile framework to address this question  
is provided by modern (functional) renormalisation group methods, based on 
the infinitesimal integrating-out of momentum degrees of freedom from a path 
integral representation of the theory \`a la Wilson \cite{Wilson:1973jj,Polchinski:1983gv,Wetterich:1992yh,Morris:1993qb,BTW}.
This is achieved by adding a momentum cutoff to the action, quadratic in the propagating fields. 
In consequence, the action \eq{S} becomes a scale-dependent effective or flowing action $\Gamma_k$, 
\begin{equation}\label{Gammak}
\Gamma_k=
\int d^dx \sqrt{g}\left[
\frac{1}{16\pi G_k}\left(-R(g_{\mu\nu})+2\Lambda_k\right) +\cdots\right] +S_{k,\rm gf}+ S_{k,\rm gh} +S_{k,\rm matter}\,,
\end{equation}
which in the context of gravity contains a running gravitational coupling, a running cosmological constant $\Lambda_k$, a gauge fixing term, ghost contributions,  matter interactions, and the dots indicate possible higher derivative operators in the metric field. Upon varying the RG scale $k$, the effective action interpolates between a microscopic  theory $\Gamma_\Lambda$ at the UV scale $k=\Lambda$ (not to be confused with the running cosmological constant $\Lambda_k$) and the macroscopic quantum effective action $\Gamma$ where all fluctuations are taken into account ($k\to 0$). The variation of \eq{Gammak} with RG scale $k$ is given by an exact functional differential equation \cite{Wetterich:1992yh,Reuter:1996cp}
\vskip-2.5cm
\begin{equation}\label{FRG}
\partial_t \Gamma_{k}
= \frac12\ {\rm Tr}\
\frac1{\Gamma_{k}^{(2)}
+ R_{k}}\
\partial_t R_{k}
\hskip-1cm  
\begin{minipage}[l]{3cm}
\includegraphics[width=4cm]{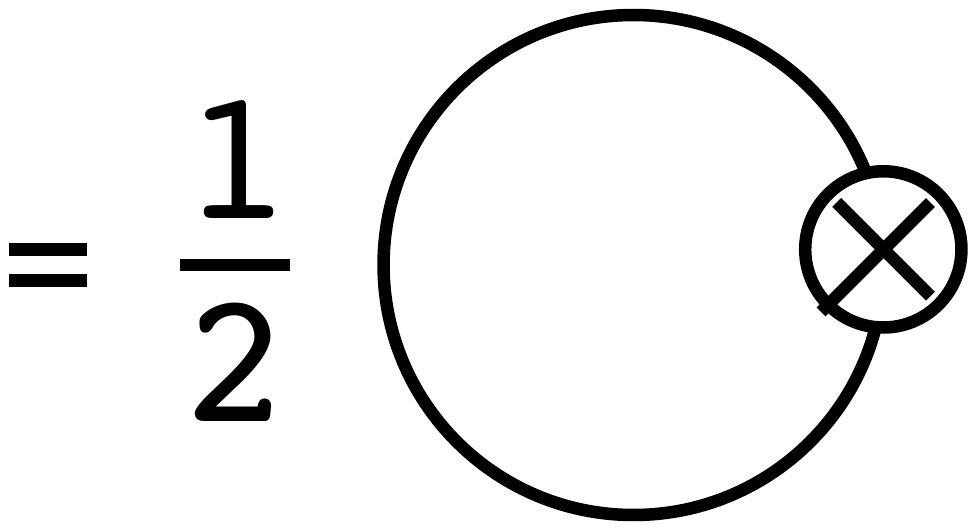}
\end{minipage}
\end{equation}
\vskip-2cm
\noindent which relates the change of the scale-dependent gravitational  action with the full field-dependent propagator of the theory (full line) and the scale-dependence $\partial_t R_k$ of the momentum cutoff (the insertion). 
Here, the trace stands for a momentum integration, and a sum over all propagating degrees of freedom $\phi=(g_{\mu\nu}$, ghosts, matter fields), which minimally contains the metric field and its ghosts. The function $R_k$
(not to be confused with the Ricci scalar) denotes the infrared momentum cutoff. As a function of (covariant) momenta $q^2$, the momentum cutoff obeys  $R_k(q^2)\to 0$ for $k^2/q^2\to 0$, $R_k(q^2)> 0$ for $q^2/k^2\to 0$, and $R_k(q^2)\to\infty$ for
$k\to\Lambda$ (for examples and plots of $R_k$, see \cite{Litim:2000ci}). The functional flow \eq{FRG} is closely linked to other exact functional differential equations such as the Callan-Symanzik equation  \cite{Symanzik:1970rt} in the limit $R_k\to k^2$, and the Wilson-Polchinski equation by means of a Legendre transformation \cite{Polchinski:1983gv,BTW,Litim:2008tt}. In a weak coupling expansion, \eq{FRG} reproduces standard perturbation theory to all loop orders \cite{Litim:2001ky,Litim:2002xm}. 
The strength of the formalism is that it is not bounded to the weak-coupling regime, and systematic approximations -- such as the derivative expansion, vertex expansions, or mixtures thereof -- are available to access domains with strong coupling and/or strong correlation. Systematic uncertainties can be assessed \cite{Litim:2010tt}, and  the stability and convergence of approximations is enhanced by optimisation techniques, see \cite{Litim:2000ci,Litim:2001fd,Litim:2001up,Litim:2002cf,Pawlowski:2005xe}.

By construction, the flow \eq{FRG} is both ultraviolet and infrared finite. 
Together with the boundary condition $\Gamma_\Lambda$  it may serve 
as a definition for the theory. In renormalisable theories, the continuum limit is performed 
by removing the scale $\Lambda$, $1/\Lambda\to 0$, and the functional $\Gamma_\Lambda\to \Gamma_*$ approaches a fixed point action at
short distances (see the proposal in~\cite{Manrique:2008zw} for a construction in gravity). 
In perturbatively renormalisable theories,
$\Gamma_*$ is "trivial" and given mainly by the classical action. In perturbatively 
non-renormalisable theories, the existence (or
non-existence) of $\Gamma_*$ has to be studied on a case-by-case basis.  
A viable fixed point action $\Gamma_*$ in quantum gravity
should at least contain those diffeomorphism invariant operators which display
relevant or marginal scaling in the vicinity of the UV fixed point.
A fixed point action qualifies as fundamental if renormalisation group trajectories 
$k\to \Gamma_k$ emanating from its vicinity
connect with the correct long-distance behaviour for $k\to 0$ and stay well-defined (finite, no poles) at all scales \cite{Weinberg}. 

The operator trace in \eq{FRG} is evaluated by using 
flat or curved backgrounds together with heat kernel techniques or plain momentum integration. 
To ensure diffeomorphism symmetry within
this set-up,  the background field formalism is used by adding a
non-propagating background field $\bar g_{\mu\nu}$
\cite{Reuter:1996cp,Litim:1998nf,Freire:2000bq,Litim:2002hj,Pawlowski:2005xe}. This way, the
extended effective action $\Gamma_k[g_{\mu\nu},\bar g_{\mu\nu}]$ becomes
gauge-invariant under the combined symmetry transformations of the physical
and the background field. A second benefit of this is that the background
field can be used to construct a covariant Laplacean $-\bar D^2$ (or similar)
to define a mode cutoff at the RG momentum scale $k^2=-\bar D^2$. This implies that
the mode cutoff $R_k$ will depend on the background fields, which is controlled by an 
equation similar to the flow equation itself \cite{Litim:2002hj,Folkerts:2011jz}. The background
field is then eliminated from the final equations by identifying it with the
physical mean field. This procedure dynamically readjusts the background field and 
implements the requirements of background independence for quantum gravity. 
An alternative technique which employs  a bi-metric approximation 
has been put forward in \cite{Manrique:2009uh}\cite{Manrique:2010mq}\cite{Manrique:2010am}. 
For a general evaluation of the different implementations of  a momentum cutoff  in gauge 
theories, see \cite{Litim:1998nf}, \cite{Litim:2002hj} and \cite{Pawlowski:2005xe}.

\section{Fixed points of quantum gravity}\label{FPs}

In this section, I give a brief summary of fixed points found so far, and refer to \cite{Litim:2008tt} for a more detailed overview of result prior to 2008. The search for fixed points in quantum gravity starts by restricting the running effective action $\Gamma_k$ to a finite set 
of operators ${\cal O}_i(\phi)$ with running couplings $g_i$,
\begin{equation}
\Gamma_k=\sum_i\,g_{i}(k)\, {\cal O}_i(\phi)\,,
\end{equation}
including eg.~the Ricci scalar $\int \sqrt{g}R$ and the cosmological constant $\int \sqrt{g}$. The flow equations for the couplings $g_i$ are obtained from \eq{FRG} by projection onto some subspace of operators. Convergence and stability of approximations is checked by increasing the number of operators retained, and  further improved through optimised choices of the momentum cutoff \cite{Litim:2000ci,Litim:2002cf}.  

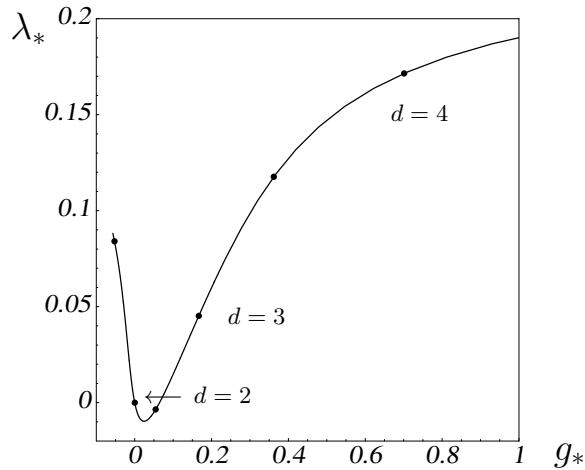
\begin{figure}[t]
\begin{center}
\unitlength0.001\hsize
\begin{picture}(400,420)
\epsfig{file=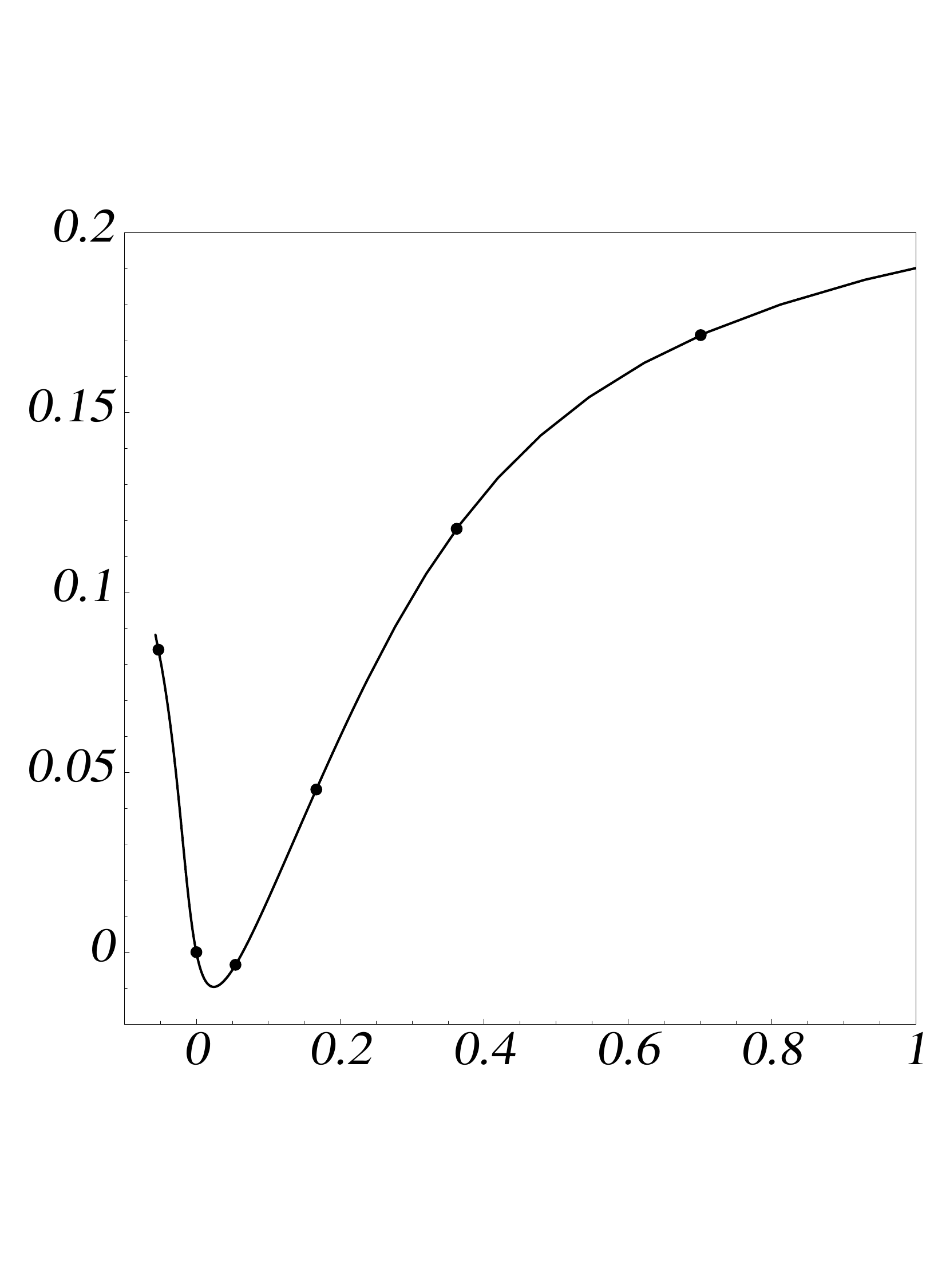,scale=0.4}
\put(20,90){\Large $g_*$}
\put(-100,340){$d=4$}
\put(-220,190){$d=3$}
\put(-282,132){$\longleftarrow\,d=2$}
\put(-380,400){\Large $\lambda_*$}
\end{picture}
\vskip-1cm
\caption{\label{Ddep} Dependence of the UV fixed point on space-time dimensionality.}
\vskip-2cm
\end{center}
\end{figure}

The first set of flow equations which carry a non-trivial UV fixed point has been derived in \cite{Reuter:1996cp} for the Einstein-Hilbert theory in Feynman gauge. As a function of the number of space-time dimensions, the flow equation reproduces the well-known fixed point in $d=2+\epsilon$ dimensions \cite{Gastmans:1977ad,Christensen:1978sc,Weinberg,Aida:1996zn} see Fig.~\ref{Ddep}. More importantly, the equations display a non-trivial UV fixed point in the four-dimensional theory \cite{Reuter:1996cp,Souma:1999at}. Both the Ricci scalar and the cosmological constant term are relevant operators at the fixed point. Their scaling is strongly correlated leading to a complex conjugate pair of universal scaling exponents. This result has subsequently been confirmed within a more general background field gauge by means of a trace-less transverse decomposition of the metric field \cite{Lauscher:2001ya}. 

\begin{figure}[t]
\begin{center}
  \unitlength0.001\hsize
\begin{picture}(500,900)
\put(0,690){\epsfig{file=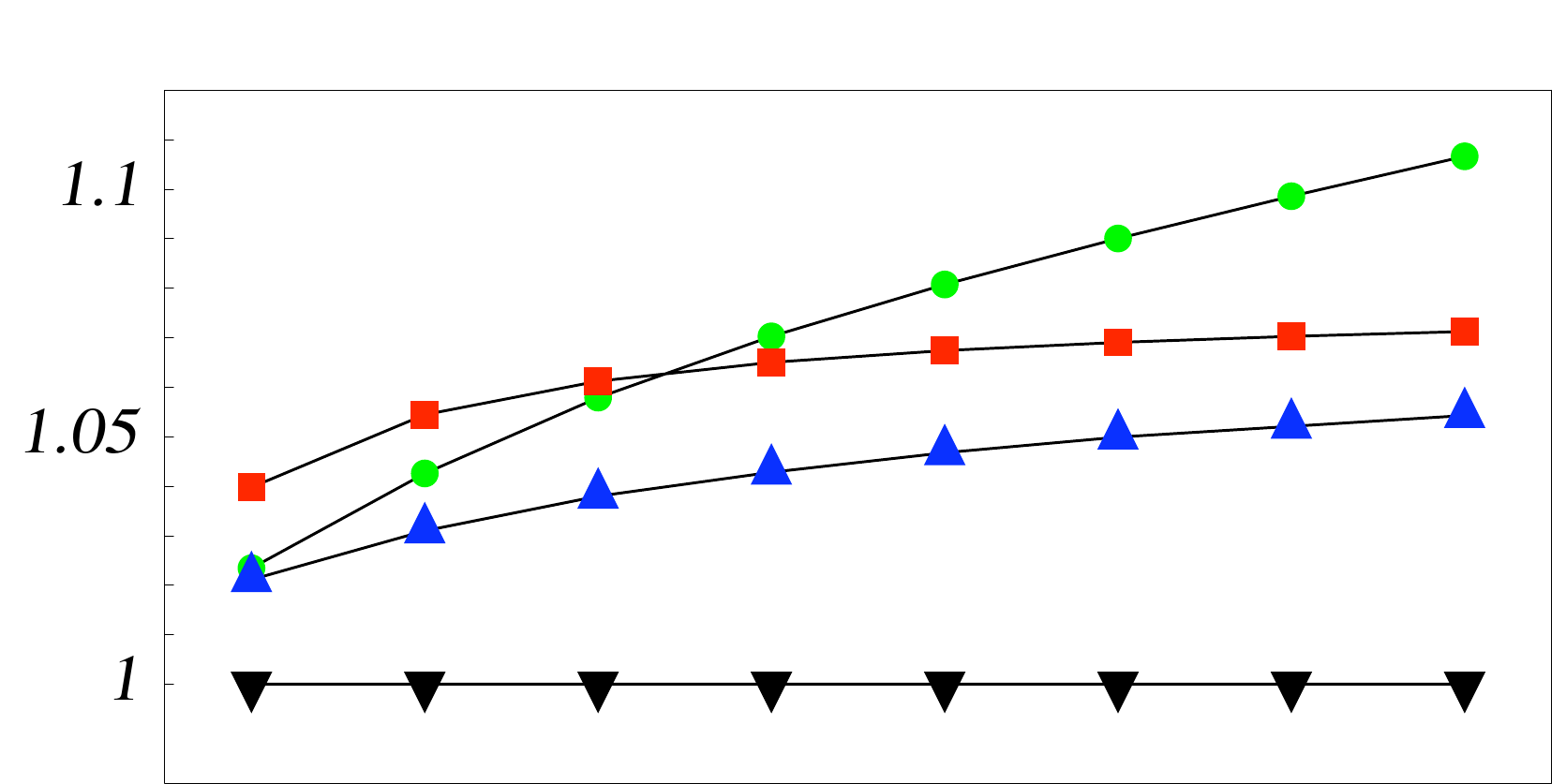,scale=0.45}}
\put(0,480){\epsfig{file=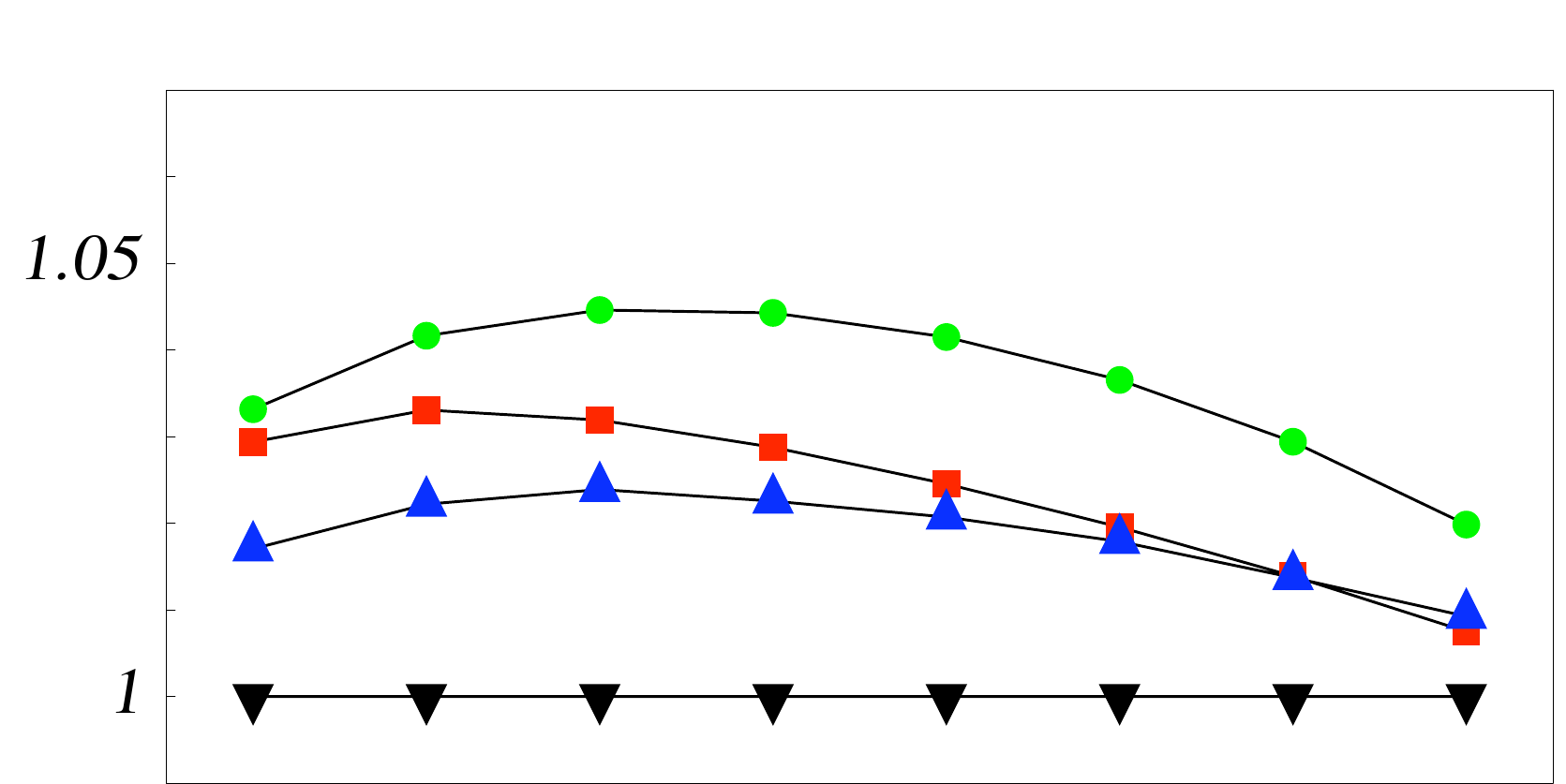,scale=0.451}}
\put(0,270){\epsfig{file=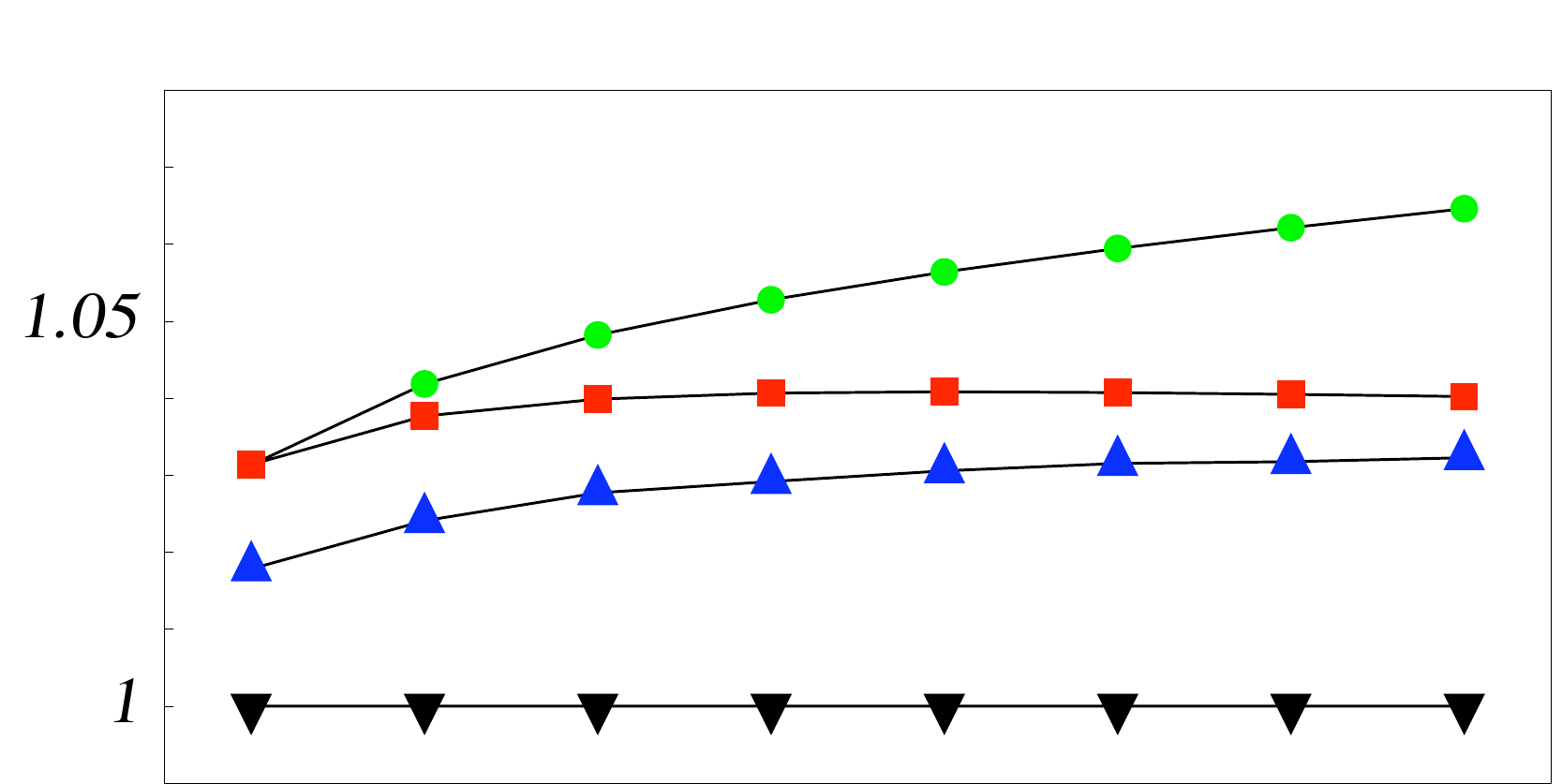,scale=0.45}}
\put(-33,30){\epsfig{file=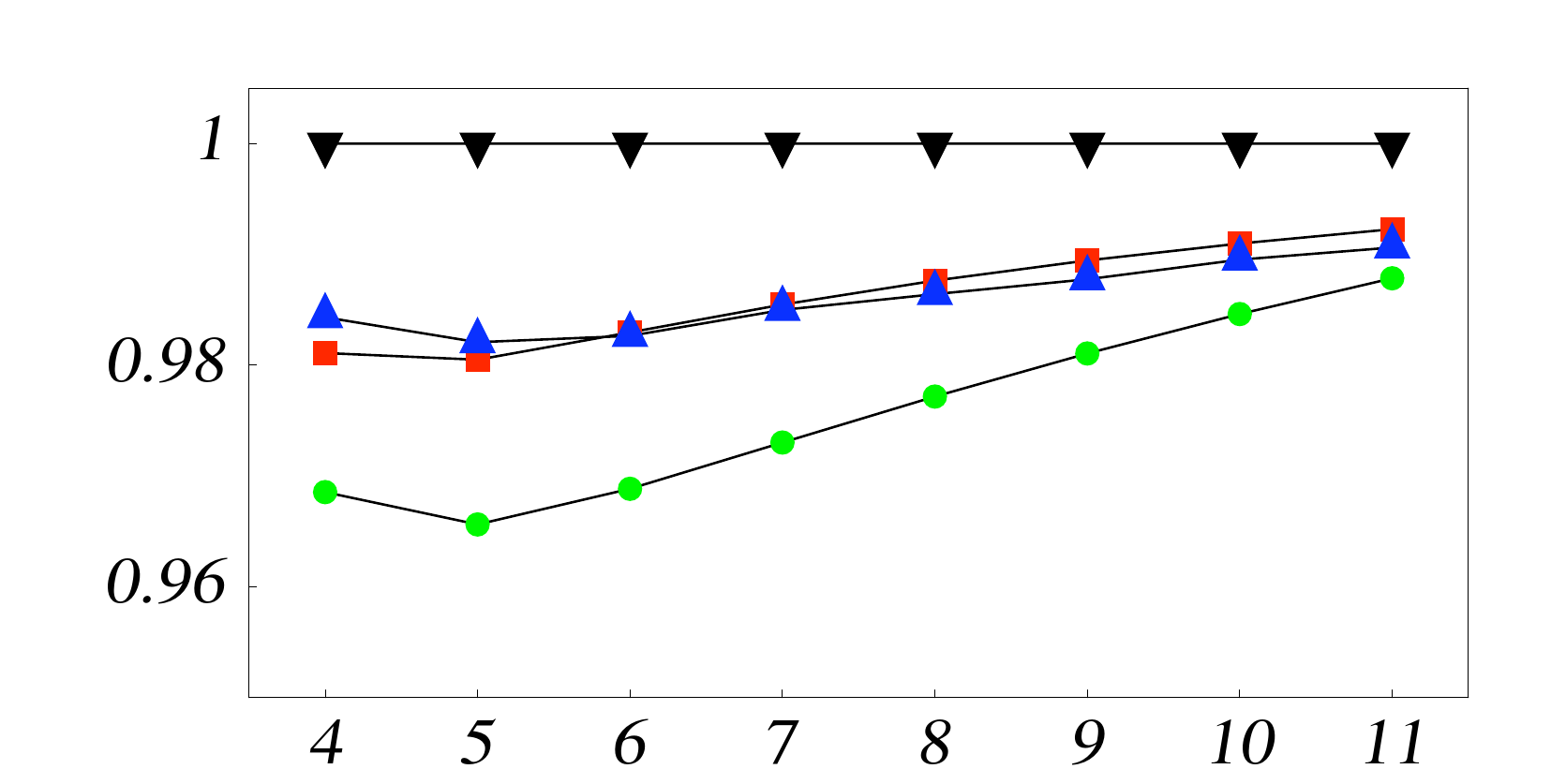,scale=0.512}}
\put(60,850){a)$\ \ \theta'/\theta'_{\rm opt}$}
\put(60,640){b)$\ \ \theta''/\theta''_{\rm opt}$}
\put(60,420){c)$\ \ |\theta|/|\theta|_{\rm opt}$}
\put(60,80){d)$\ \ \tau/\tau_{\rm opt}$}
\put(220,0){\large \bf $d$}
\end{picture}
\caption{\label{Vergleich}Ultraviolet fixed point in the $d$-dimensional Einstein-Hilbert theory. Comparison of universal eigenvalues $\theta'$, $\theta''$,
  $|\theta|$ and the invariant $\tau=\lambda_*(g_*)^{2/(d-2)}$ for different Wilsonian momentum cutoffs and various dimensions, normalised to the
  result for $R_{\rm opt}$ ($R_{\rm mexp}$ \textcolor{green}{$\bullet$},
  $R_{\rm exp}$ \textcolor{red}{$\blacksquare$}, $R_{\rm mod}$
  \textcolor{blue}{$\blacktriangle$}, $R_{\rm opt}$ $\blacktriangledown$); from \cite{Fischer:2006fz}.}
\end{center}
\end{figure}
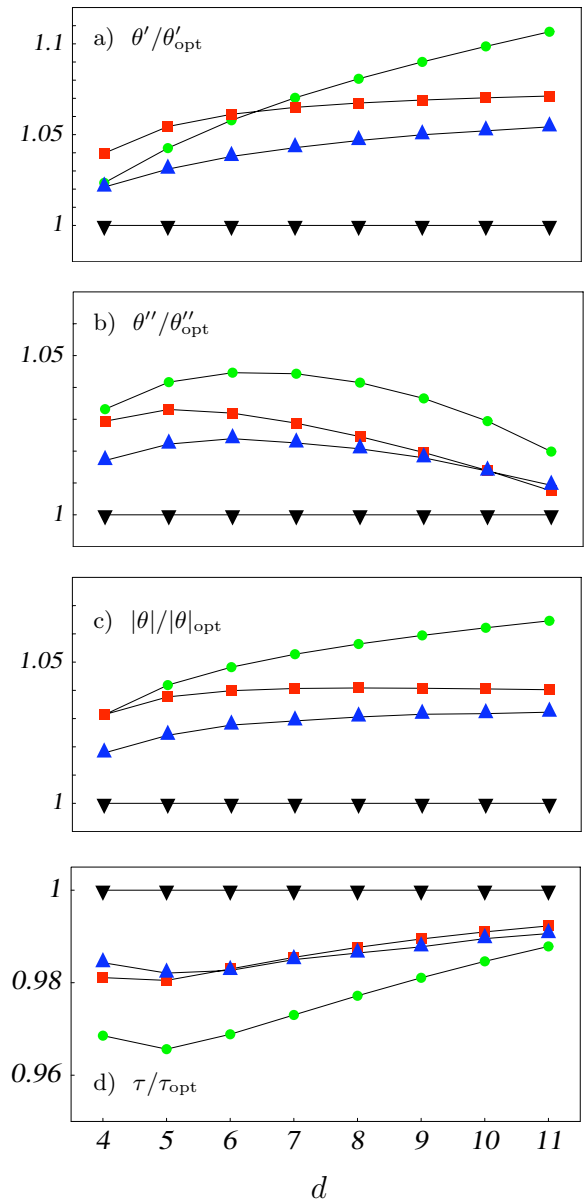

\begin{figure}
  \unitlength0.001\hsize
\begin{center}
\begin{picture}(600,540)
\put(400,500){\large$g$}
\put(400,460){\large$\lambda$}
\put(400,380){\large$\eta$}
\put(550,250){\large $\ln k/\Lambda_T$}
\put(120,380){ IR}
\put(480,380){ UV}
\includegraphics[scale=.5]{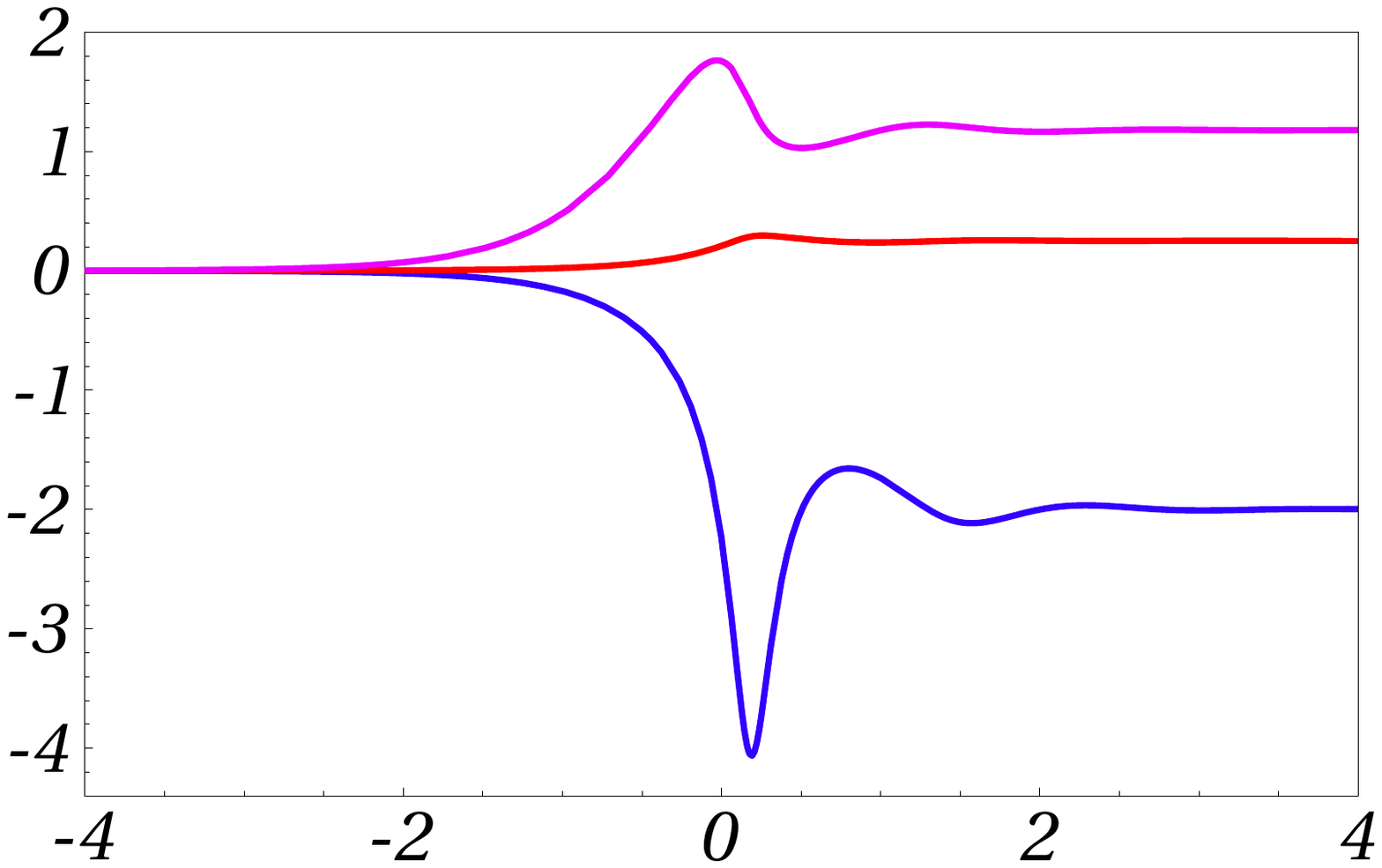}
\end{picture}
\vskip-4cm
\caption{\label{running}Running couplings along the separatrix in four dimensions; from \cite{Litim:2003vp}.}
\end{center}
\end{figure}

A complete analytical understanding of the Einstein-Hilbert approximation has been achieved in \cite{Litim:2003vp}, leading to analytical flow equations and closed expressions for the (unique) UV fixed point and its universal eigenvalues. Key for this was the use of (optimised) momentum cutoffs which allow analytical access and improve convergence and stability of results \cite{Litim:2001up,Litim:2000ci}. 
Fig.~\ref{Vergleich}  shows 
the universal scaling exponents $\theta=\theta'+i\theta''$ at the UV fixed point in four dimensions (first column) for various (optimised) choices of the momentum cutoff \cite{Fischer:2006fz,Litim:2001up}. Also on display is the invariant $\tau=\lambda_*(g_*)^{2/(d-2)}$. The variations in either of these are small. The dependence on the gravitational gauge fixing parameter $\alpha$ is moderate and controlled by an independent fixed point in the gauge fixing sector $ie.$ Landau- de Witt gauge ($\alpha=0$) \cite{Litim:1998qi}. Fig.~\ref{running} shows the RG running of couplings along the separatrix connecting the non-trivial UV fixed point with the Gaussian fixed point in the infrared. At the scale $k\approx\Lambda_T$, the RG flow displays a cross-over from perturbative IR scaling to fixed point scaling where the anomalous dimension grows large. The dynamical scale $\Lambda_T$ is of the order of the fundamental Planck scale, and a consequence of the UV fixed point. In gravity, the scale $\Lambda_T$ plays a r\^ole analogous to that  of $\Lambda_{\rm QCD}$ in quantum chromodynamics.

In a formidable tour-de-force computation, the space of operators has been extended to include $\int \sqrt{g}R^2$ interactions \cite{Lauscher:2001rz,Lauscher:2002mb}. The perturbatively marginal $R^2$ coupling turned out to become relevant at a UV fixed point. Elsewise, it added only minor corrections to the fixed point in the Einstein-Hilbert theory. With \cite{Litim:2003vp,Litim:2001up}, these computations became feasible analytically and have been implemented for operators $\int \sqrt{g}\,f_k(R)$ which contain a general function of the Ricci scalar in \cite{Codello:2007bd,Codello:2008vh} and \cite{Machado:2007ea}. A non-trivial UV fixed point has been found by Taylor-expanding $f_k(R)$ to high orders in the Ricci scalar with results known up to eighth order in the eigenvalues \cite{Codello:2008vh}  and up to tenth order in the fixed point  \cite{Bonanno:2010bt}. Additionally, the fixed is remarkably stable. The universal eigenvalues converge rapidly, confirming all previous results. Most importantly, it turned out that operators $\int\sqrt{g}R^n$ from $n\ge 3$ onwards become {\it irrelevant} at the UV fixed point \cite{Codello:2007bd}, a first indicator that an asymptotically safe UV fixed point of the fundamental theory may in fact have a finite number of relevant operators. This is most welcome as otherwise a fixed point with infinitely many relevant couplings is likely to spoil the predictive power. 

All fourth order derivative operators have been included both at the one-loop order and beyond. The re-evaluation of perturbative one-loop results by means of a functional flow emphasized the significance of eg.~quadratic divergences which normally are suppressed within dimensional regularisation \cite{Codello:2006in,Niedermaier:2009zz,Niedermaier:2010zz}. Two types of UV fixed points are found. The first one is non-trivial in all couplings \cite{Benedetti:2009rx,Benedetti:2009gn} with three relevant directions at the fixed point. The second one is perturbative in the Weyl coupling  \cite{Codello:2006in,Niedermaier:2009zz} but non-trivial in Newton's coupling, with two relevant and two marginally relevant directions. In either case Newton's coupling takes a non-trivial fixed point in the UV. 
These result hence indicate that higher derivative operators are compatible with gravity developing a non-trivial fixed point. Once this is achieved, the difference between the Einstein-Hilbert approximation and the fourth order approximation is qualitatively small. 
In the above, the RG running of the ghost sector is expected to be subleading. This has recently been confirmed in  \cite{Groh:2010ta} and in \cite{Eichhorn:2010tb}, and for the scalar curvature-ghost coupling in \cite{Eichhorn:2009ah}.

An important extension deals with the inclusion of matter couplings. In eg.~Yang-Mills theory, it is well-known that the property of asymptotic freedom can be spoiled by too many species of fermions. In the same vein it has to be checked whether the UV fixed point of gravity remains stable under the inclusion matter, and vice versa. By now, this question has been analysed for minimally coupled scalar matter  \cite{Percacci:2002ie}, non-minimally coupled scalar fields \cite{Narain:2009gb,Narain:2009fy}, and generic free matter coupled to gravity \cite{Codello:2006in,Percacci:2003jz}. 
These findings are consistent with earlier results using different techniques in the limit of many matter fields \cite{Tomboulis:1977jk}\cite{Smolin:1981rm}.
The conclusion is that the gravitational fixed point generically persists under the inclusion of matter (including the Standard Model and its main extensions) unless a significant imbalance between bosonic and fermionic matter fields is chosen. Also, non-minimally coupled scalar matter leads to slight deformation of the gravitational fixed point with matter achieving a weakly coupled regime, the so-called Gaussian matter fixed point. Fermions have been included recently in \cite{Zanusso:2009bs}, as well as gravitational Yukawa systems \cite{Vacca:2010mj}. In the latter case, the coupled system displays a non-trivial UV fixed point both in the Yukawa as well as in the gravitational sector.  This new fixed point may become of great interest as a stabilizer for the Standard Model Higgs. 
On the one-loop level, the interplay between a Standard Model Higgs and asymptotically safe gravity has been addressed in \cite{Shaposhnikov:2009pv}, detailing conditions under which no new physics is required to bridge the energy range from the electroweak scale all the way up to \eq{Planck}. In a similar spirit, it has been argued that the consistency of Higgs inflation models is enhanced provided that gravity becomes asymptotically safe  \cite{Atkins:2010yg}.

The impact of gravitational fluctuations on gauge theories has seen re-newed interest initiated by a one-loop study within effective theory \cite{Robinson:2005fj}. Asymptotic freedom is sustained by gravity leading to a vanishing \cite{Pietrykowski:2006xy,Toms:2007sk,Ebert:2007gf,Toms:2008dq,Folkerts:2011jz}
or non-vanishing one-loop correction \cite{Robinson:2005fj,Tang:2008ah,Daum:2010bc,Toms:2010vy,Folkerts:2011jz}, depending on the regularisation.
In the present framework \eq{Gammak}, \eq{FRG}, the gravity-induced corrections have been studied in  \cite{Daum:2010bc} and in \cite{Folkerts:2011jz}. In either case the sign of the graviton contributions is fixed and asymptotic freedom of Yang-Mills theory persists in the presence of gravitational fluctuations, also including a cosmological constant. Furthermore, \cite{Folkerts:2011jz} evaluates  generic regularisations and clarifies the non-universal nature of the one-loop coefficient, also covering generic field-theory-based UV scenarios for gravity including  asymptotically safe gravity and gravitational shielding. 
An interesting consequence of a non-trivial gravitational contribution to the running of an abelian charge is the appearance of a combined UV fixed point  in the $U(1)$-gravity system providing a mechanism to calculate the fine structure constant \cite{Harst:2011zx}.

In three dimensions, a fixed point study has been reported within topological massive gravity which includes a Chern-Simons (CS) term in addition to Newton's coupling and the cosmological constant \cite{Percacci:2010yk}. On the one-loop level, the CS term has a vanishing beta function and the gravitational sector displays both a Gaussian and a nontrivial fixed point for positive CS coupling which is qualitatively in accord with the fixed point in Fig.~\ref{Ddep} (no CS term). 

An interesting structural link has been noted between the gravitational fixed point \cite{Litim:2003vp} (Ricci scalar, no cosmological constant) and a UV fixed point in non-linear sigma models \cite{Codello:2008qq}. In either system the beta-function of the relevant coupling displays a UV fixed point whose coordinates are non-universal. However, the systems display the same universal scaling, with identical exponent $\theta=2d(d-2)/(d+2)$ for all  $d\ge 2$ dimensions. This coincidence 
highlights that non-linear sigma models have a non-trivial UV dynamics on their own, possibly quite similar to gravity itself, which is worth of being explored further \cite{Percacci:2009dt}.  Fixed points in more complex (gauged and non-gauged) non-linear sigma models have recently been obtained in \cite{Percacci:2009fh,Fabbrichesi:2010xy}.

It would be very helpful to understand which mechanism implies the gravitational fixed point by eg.~restricting the dynamical content. In fact, a two Killing vector reduction of four-dimensional gravity (with scalar matter and photons) has been shown to be asymptotically safe  \cite{Niedermaier:2003fz}. This topic has also been studied by reducing gravity to the dynamics of its conformal sector   \cite{Reuter:2008wj,Reuter:2008qx}  including higher derivative terms and effects from the conformal anomaly \cite{Machado:2009ph}. Although a UV fixed point is visible in most approximations, its presence depends more strongly on eg.~matter fields,
meaning that more work is required to settle this question for pure gravity.

 Implications of a gravitational fixed point have been studied in the context of black holes, astrophysics and infrared gravity, and cosmology.  
The RG induced running $G_N\to G(r)$  of Newton's coupling has interesting implications for the existence of RG improved black hole space-times \cite{Bonanno:2000ep} as well as for their dynamics \cite{Bonanno:2006eu}. If gravity weakens according to its asymptotically safe RG flow, results point towards the existence of a smallest black hole with critical mass $M_c$ of the order of $M_{\rm Pl}$. This pattern persists for Schwarzschild black holes in higher dimensions \cite{Falls:2010he}, rotating black holes \cite{Reuter:2010xb}, and black holes within higher derivative gravity \cite{Cai:2010zh}, showing that the existence of smallest black holes is a generic and stable prediction of this framework.

Low energy implications of the RG running of couplings have equally been studied, including applications to galaxy rotation curves \cite{Reuter:2007de,Bonanno:2009nj}.  Recent studies have re-evaluated gravitational collapse \cite{Casadio:2010fw} exploring the one-loop running couplings, and the RG induced non-local low energy corrections to the gravitational effective action \cite{Satz:2010uu}. It has also been conjectured that gravity may possess an infrared fixed point. In this case, Newton's coupling grows large at large distances and may contribute to an accelerated expansion of the universe, a scenario which can be tested against  cosmological data such as Type Ia supernov\ae\ \cite{Bonanno:2001hi,Bentivegna:2003rr,Reuter:2005kb}.  

The intriguing idea that an asymptotically safe UV fixed point may control the beginning of the universe has been implemented in Einstein gravity  with an ideal fluid \cite{Bonanno:2001xi,Bonanno:2002zb}, and in the context of $f(R)$-theories of gravity \cite{Weinberg:2009wa, Tye:2010an,Bonanno:2010bt,Contillo:2010ju}. The RG scale parameter is linked with cosmological time and therefore leads to RG improved cosmological equations \cite{Bonanno:2001xi,Hindmarsh:2011hx,Koch:2010nn}, which may even generate entropy  \cite{Bonanno:2007wg,Bonanno:2010mk}. 
The framework put forward in \cite{Hindmarsh:2011hx}  includes scalar fields and exploits the Bianchy identity, leading to general conditions of existence for cosmological fixed points (with cosmic time) whose solutions include eg.~inflating UV safe fixed points of gravity coupled to a scalar field.

The proposal that Lorentz symmetry can be broken by gravity on a fundamental level has been put forward in    \cite{Horava:2009uw} where space and time scale differently in the UV with a relative dynamical exponent $z\neq 1$. Provided $z=3$ (in four dimensions) gravity is power-counting renormalisable in perturbation theory and Lorentz symmetry should emerge as a low energy phenomenon   \cite{Horava:2011gd}. Interestingly, Horava-Lifshitz gravity with $z\neq3$ also requires an asymptotically safe fixed point for  gravity with an anomalous dimension $\eta\neq 0$ compensating for power-counting non-renormalisability.

The RG scaling in the vicinity of an asymptotically safe fixed point implies scale invariance. On the other hand, it is often assumed that a quantum theory of gravity should, in one way or another, induce a minimal length. Different aspects of this question and interrelations with other approaches to quantum gravity have been addressed in \cite{Reuter:2005bb,Reuter:2006zq}, and more recently in \cite{Basu:2010nf,Percacci:2010af} and \cite{Calmet:2010tx}. The findings indicate that the fixed point {\it implicitly} induces a notion of minimal length, related to the RG trajectory and fundamental scale where gravity crosses over from classical to fixed point scaling. It has also been argued that RG corrections to gravity can modify the dispersion relation of massive particles leading to indirect bounds on the one-loop coefficients of the gravitational beta-functions \cite{Girelli:2006sc} .

Finally, it is worth noting that standard quantum field theories (without gravity) may become asymptotically safe in their own right. This has first been exemplified for perturbatively non-renormalisable Gross-Neveu models in three dimensions with functional RG methods \cite{Gawedzki:1985ed} and in the limit of many fermion flavours \cite{deCalan:1991km}. In four dimensions, this possibility has been explored for variants of the Standard Model Higgs \cite{Gies:2003dp}, for Yukawa-type interactions \cite{Gies:2009hq,Gies:2009sv}, and strongly coupled gauge theories \cite{Pica:2010xq}.  It has also been conjectured that the electroweak theory without a dynamical Higgs field could become asymptotically safe \cite{Calmet:2010ze}, based on structural similarities with quantum general relativity.

In summary, there is an increasing amount of evidence for the existence of non-trivial UV fixed points in four-dimensional quantum field theories including gravity and matter.

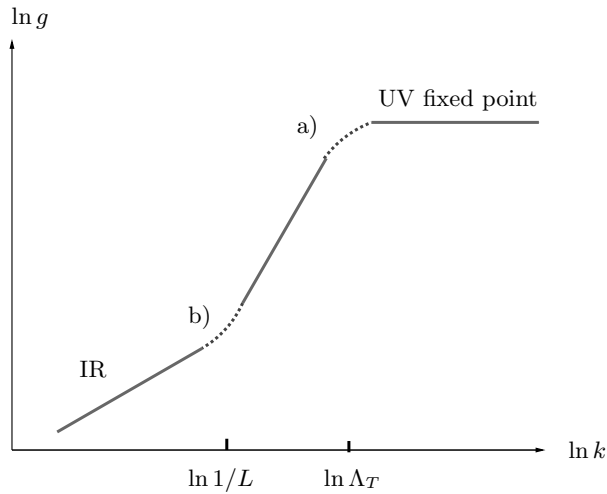
\begin{figure}
\begin{picture}(1000,20)
\end{picture}
\unitlength0.001\hsize
\includegraphics[width=.4\hsize]{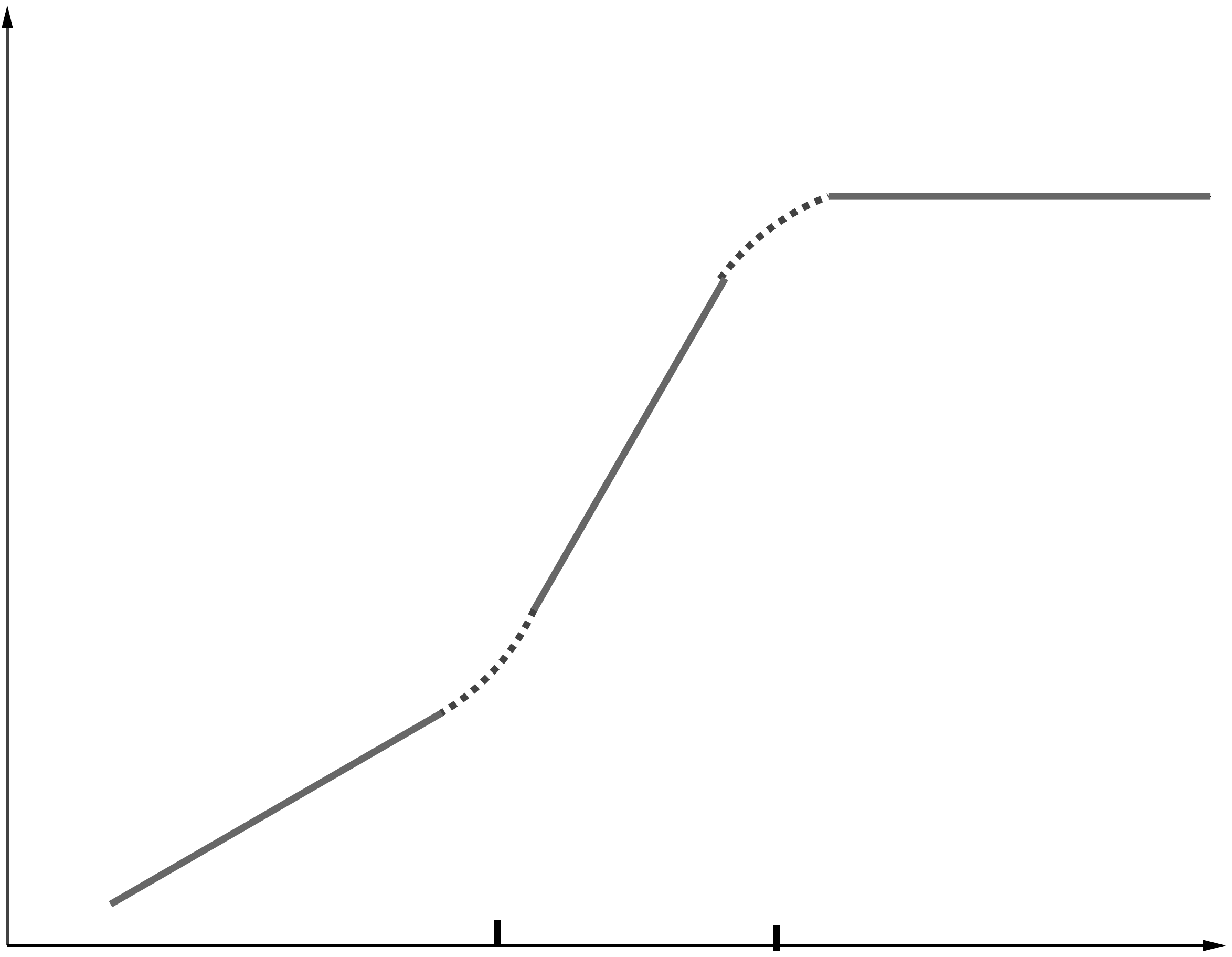}
\begin{picture}(1000,5)
\put(300,340){$\ln g$}
\put(570,280){UV fixed point}
\put(350,80){IR}
\put(510,260){{a)}}
\put(430,120){{b)}}
\put(710,20){$\ln k$}
\put(530,0){$\ln \Lambda_T$}
\put(430,0){$\ln 1/L$}
\end{picture}
\caption{Log-log plot (schematic) of the dimensionless running gravitational coupling in a scenario
  with large extra dimensions of size $\sim L$ and a dynamical Planck scale
  $\Lambda_T$ with $\Lambda_T\,L\gg 1$. The fixed point behaviour in the deep ultraviolet
  enforces a softening of gravity.
   a) At the scale $k\approx \Lambda_T$ the coupling displays a cross-over from fixed point scaling to classical scaling in the higher-dimensional theory.
  b) At the compactification scale  $k\approx 1/L$, the $n$ compactified spatial dimensions are no longer available for gravity to propagate in, and the running coupling displays a cross-over from $(4+n)$-dimensional to $4$-dimensional scaling.}
\label{RunningG}
\end{figure}

\section{Extra dimensions}\label{Extra}

Next I turn to fixed points of quantum gravity in
$d=4+n$ dimensions where $n$ denotes the number of extra dimensions  \cite{Litim:2003vp,Fischer:2006fz,Fischer:2006at}, and potential signatures thereof in models with a low quantum gravity scale \cite{Litim:2007iu,Litim:2007ee,Hewett:2007st,Gerwick:2011jw,Gerwick:2010kq}.
There are several motivations for this. Firstly, particle physics models where gravity lives in a higher-dimensional space-time  have raised enormous interest  by allowing the fundamental Planck scale to be as low as the electroweak scale \cite{ArkaniHamed:1998rs,Antoniadis:1998ig,Randall:1999ee}, 
\begin{equation}
M_*\approx  1 - 10\ {\rm TeV}\,.
\label{lowPlanck}
\end{equation}
This way, the notorious hierarchy problem of the Standard Model (SM) is circumnavigated. If realized in Nature, this opens the exciting possibility that Planck scale physics becomes accessible to experiment via eg.~high-energetic particle collisions at the LHC.
The main point of these models is that the four-dimensional Planck scale \eq{Planck} is no longer fundamental but a derived quantity, related as $M^2_{\rm Pl}\sim M^2_{*} (M_*\,L)^{n}$ to the Planck scale \eq{lowPlanck} of the higher-dimensional theory and the size of the compact extra dimensions $L$.

Secondly, the critical dimension of gravity -- the dimension where the gravitational
coupling has vanishing canonical mass dimension -- is two. Hence, for any dimension
above the critical one, the canonical dimension of Newton's coupling is negative. 
From a renormalisation group point of view, this means that four dimensions appear 
not to be special. Continuity in the dimension suggests that an ultraviolet fixed point,
if it exists in four dimensions, should persist towards higher dimensions. Furthermore, 
gravity has $d(d-3)/2$ propagating degrees of freedom
which rapidly increases with increasing dimensionality.
It is important to understand whether these additional degrees of freedom spoil or support the UV fixed point 
detected in the four-dimensional theory. The local structure of quantum fluctuations,
and hence local renormalisation group properties of a quantum theory of
gravity, are qualitatively similar for all dimensions above the critical one,
modulo topological effects for specific dimensions.  
Therefore one should expect to find similarities in the ultraviolet behaviour of gravity in four
and higher dimensions.  
 
\begin{figure*}[t]
\begin{center}
\unitlength0.001\hsize
\begin{picture}(1000,280)
\includegraphics[width=.9\textwidth]{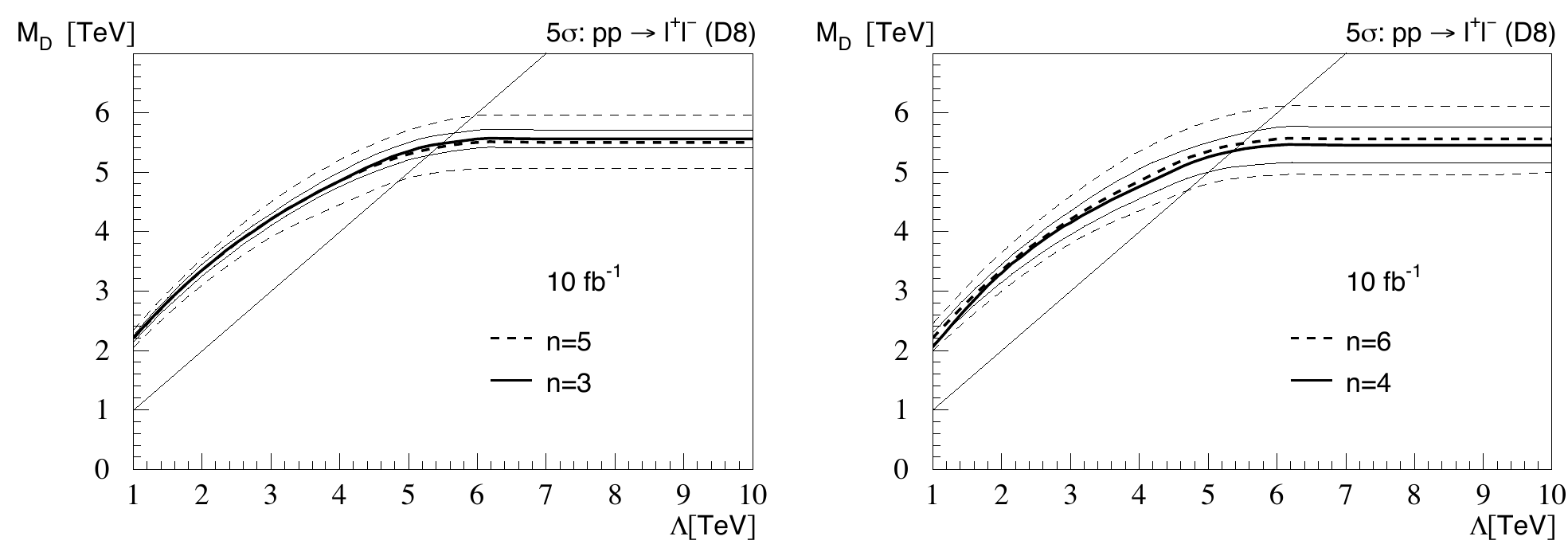}
\end{picture}
\caption{The $5\sigma$ discovery contours in $M_*$ at the LHC as a
  function of a cutoff $\Lambda$ on $E_{\rm parton}$ for an assumed integrated
  luminosity of $10\,{\rm fb}^{-1}$. Note that the limit $\Lambda\to\infty$ can be 
  performed, and the leveling-off
  at $M_*\approx \Lambda$ reflects the underlying fixed
  point (thin dashed lines show a $\pm$10\% variation about the transition scale $\Lambda_T=M_*$); 
  from \cite{Litim:2007iu}.}
\label{fig:discovery}
\end{center}
\end{figure*}

This expectation has been confirmed in \cite{Litim:2003vp} for  $d$-dimensional Einstein-Hilbert gravity. The stability of the result has subsequently been tested 
through an extended fixed points search in higher-dimensional gravity for general cutoffs $R_k$ \cite{Fischer:2006fz}, also
probing the stability of the fixed point against variations of the gauge fixing \cite{Fischer:2006at}, and the results are displayed in 
Fig.~\ref{Vergleich}. The cutoff variations are very moderate, and smaller than the variation with gauge fixing parameter.  The structural stability of the fixed point also strengthens the findings in the four-dimensional case.

Thus, the following picture emerges: In the setting with compact extra dimensions the running dimensionless coupling $g$ displays a cross-over from fixed point scaling to $d$-dimensional classical scaling at the cross-over scale $\Lambda_T$ of the order of $M_*$, see Fig.~\ref{RunningG}a). By construction, this has to happen at energies way above $1/L$. At lower energies, close to the compactification scale  $\sim 1/L$, the $n$ compactified spatial dimensions are no longer available for gravity to propagate in, and the running coupling displays a cross-over from $(4+n)$-dimensional to $4$-dimensional scaling see Fig.~\ref{RunningG}b). 
High energetic particle colliders such as the LHC are sensitive to the electroweak energy scale, and hence to the regime of Fig.~\ref{RunningG}a), provided that the fundamental Planck scale is of the order of \eq{lowPlanck}. A generic prediction of large extra dimensions is  a tower of massive Kaluza-Klein gravitons \cite{Giudice:1998ck,Han:1998sg,Hewett:1998sn}. In particular, the exchange of virtual gravitons in Drell-Yan processes will lead to deviations in Standard Model reference processes such as $pp \to \ell^+\ell^-$ \cite{Ellis:1991qj}.
Gravitational Drell Yan production is mediated through tree-level graviton exchange and via graviton loops. 
Within effective theory, the corresponding effective operators are ultraviolet divergent (in two or more extra dimensions) and require an UV cutoff. Within fixed point gravity, the graviton is dressed by its anomalous dimension leading to finite and cutoff independent results for cross sections \cite{Litim:2007iu}, see \cite{Gerwick:2009zx} for a comparison of different UV completions.

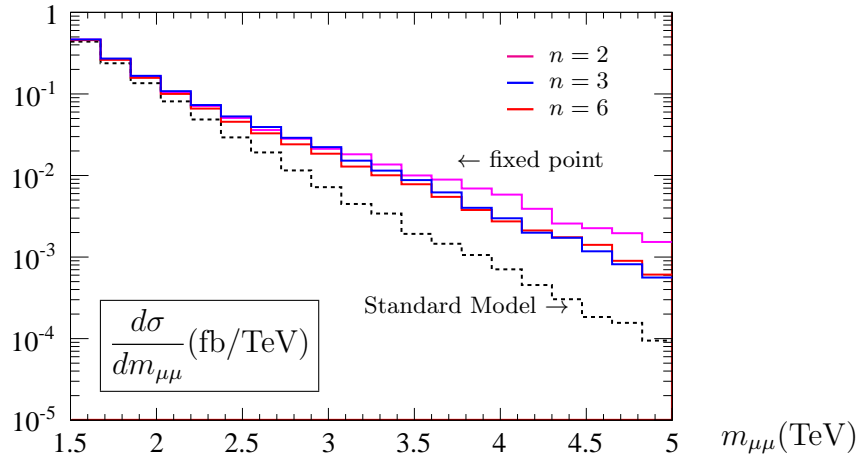
\begin{figure}[t]
\includegraphics[scale=.5]{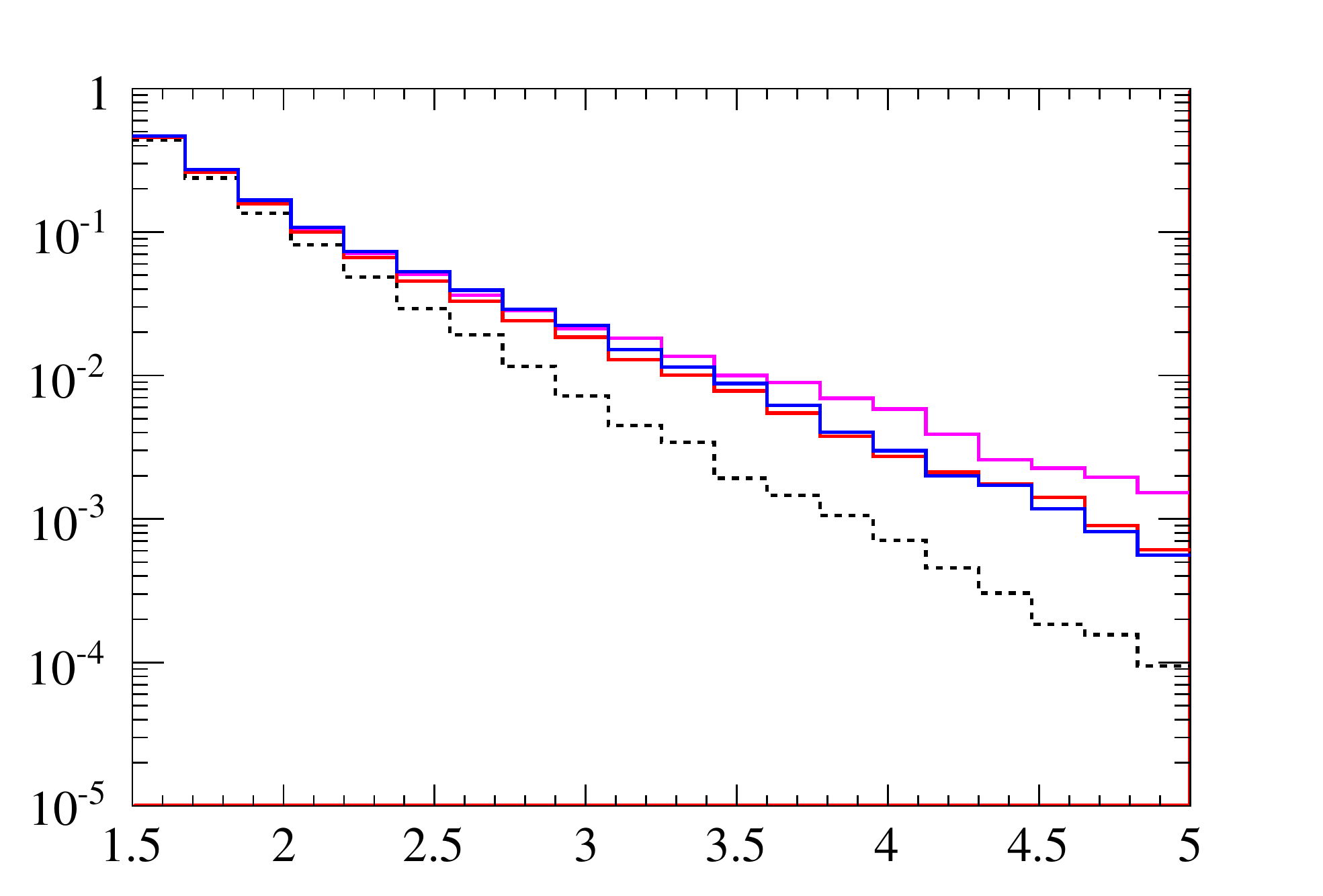} 
\put(-245,45){\framebox{\large $\dfrac{d\sigma}{dm_{\mu\mu}}{\large (\text{fb/TeV})}$}}
\put(-10,10){{\large $m_{\mu\mu}({\rm TeV})$}}
\put(-145,60){Standard Model $\rightarrow$}
\linethickness{.3mm}
\put(-91,157){\color{magenta}\line(2,0){10}}
\put(-75,155){$n=2$}
\put(-91,147){\color{blue}\line(2,0){10}}
\put(-75,145){$n=3$}
\put(-91,137){\color{red}\line(2,0){10}}
\put(-75,135){$n=6$}
\put(-110,115){$\leftarrow$ fixed point}
\caption{Differential cross-sections for di-muon production at the LHC as a function of the di-lepton invariant mass $m_{\mu\mu}$ within asymptotically safe gravity for $M_*=\Lambda_T=5$ TeV and  $n=2$ (magenta), $n=3$  (blue) and $n=6$ (red) extra dimensions, in comparison with Standard Model background (black dashed line); from~\cite{Gerwick:2011jw}.}
\label{fig:SM}
\end{figure}

Fig.~\ref{fig:discovery} displays the $5\sigma$ discovery reach to detect the fundamental scale of gravity assuming $M_*=\Lambda_T$ and an integrated luminosity of $10\,{\rm fb}^{-1}$ at the LHC. Note that the curves are given as functions of an energy cutoff $\Lambda$. In effective theory, these curves continue to grow with $\Lambda$, meaning that the cutoff cannot be removed. Here, the leveling-off is a dynamical consequence of the underlying fixed point, and the amplitude is independent of any cutoff. Fig.~\ref{fig:SM} shows the differential cross-sections for di-muon production within asymptotically safe gravity, with cross-over scale $\Lambda_T=M_*$ \cite{Gerwick:2011jw}. For all dimensions considered, the enhanced di-lepton production rate sticks out over Standard Model backgrounds, leading to the conclusion that Drell Yan production is very sensitive to the quantum gravity scale. 

Further collider signatures of fixed point gravity in warped and large extra dimensions have been addressed in \cite{Hewett:2007st} within a form-factor approximation \cite{Gerwick:2011jw}. 
Modifications to the semi-classical production of Planck-size black holes at the LHC induced by the fixed point have been analysed in \cite{Falls:2010he,Burschil:2009va}. The main new effect is an additional  suppression of the production cross section for mini-black holes. Some properties of Planck-size black holes have also been evaluated within a one-loop approximation, provided the gravitational RG flow is driven by many species of particles \cite{Calmet:2008tn}.
Graviton loop corrections  to electroweak precision observables within asymptotically safe gravity with extra dimensions have recently been obtained in \cite{Gerwick:2010kq}. 

In summary, experimental signatures for the quantisation of gravity within the asymptotic safety scenario are in reach for particle colliders, provided the fundamental scale of gravity is as low as \eq{lowPlanck}.

\section{Conclusion}\label{C}
Renormalisation group methods have become a key tool in the attempt to understand gravity at shortest, and possibly at largest, distances. It is conceivable that Planckian energies \eq{Planck} can be assessed with the help of running gravitational couplings such as \eq{gk}, replacing \eq{g} from perturbation theory. If gravity becomes asymptotically safe, its ultraviolet fixed point acts as an anchor for the underlying quantum fluctuations. The increasing amount of evidence for a gravitational fixed point with and without matter and its significance for particle physics and cosmology are very promising and certainly warrant farther reaching investigations. 

\acknowledgements{These notes are based on invited plenary talks at the conference {\it The Exact Renormalisation Group 2010} (Corfu, Sep 2010), at the workshop {\it Critical Behavior of Lattice Models in Condensed
Matter and Particle Physics} (Aspen Center for Physics, Jun 2010),  at the IV Workshop {\it Hot Topics in Cosmology} (Cargese, May 2010),  at the INT workshop {\it New Applications of the Renormalization Group Method} (Seattle, Feb 2010),  at the conference {\it Asymptotic Safety} (Perimeter Institute, Nov 2009), at the XXV Max Born Symposium {\it The Planck Scale} (Wrozlaw, Jun 2009), and on an invited talk at {\it GR19:\, the 19th International Conference on General Relativity and Gravitation} (Mexico City, Jul 2010).
I thank  Dario Benedetti, Mike Birse, Denis Comelli, Fay Dowker, Jerzy Kowalski-Glikman, Yannick Meurice, Roberto Percacci, Nikos Tetradis, Roland Triary and Shan-Wen Tsai  for their invitations and hospitality, and my collaborators and colleagues for discussions. This work was supported by the Royal Society, and by the Science and Technology Research Council [grant number ST/G000573/1].}

%********|*********|*********|*********|*********|*********|*********|****

\end{document}